\documentclass{cernrep}
\usepackage{graphicx}
\usepackage{xcolor}
\usepackage[bookmarks, colorlinks=true, linktoc=page, linkcolor=black, citecolor=black, urlcolor=blue]{hyperref}
\usepackage{enumitem}
\usepackage{ulem}
\usepackage{import}
\usepackage{fancyhdr}
\fancyhfoffset{4 mm}
\fancypagestyle{ARTTITLE}{%
\fancyhf{} 
\lhead{\small{Proceedings of the 2018 CERN--Accelerator--School course on\\
\it{Numerical Methods for Analysis, Design and Modelling of Particle Accelerators}, Thessaloniki, (Greece)}} 
\lfoot{Available online at \url{https://cas.web.cern.ch/previous-schools}}
\rfoot{\thepage\hspace*{3mm}}
 
}

\usepackage{tikz}
\usetikzlibrary{shapes,snakes}
\usetikzlibrary{arrows, decorations.markings}  
\tikzstyle{mybox} = [draw=black, fill=black!10, very thick,
  rectangle, rounded corners, inner sep=0.5cm, inner ysep=0.5cm]
\tikzstyle{fancytitle} =[draw=black, fill=white, text=black]
\tikzstyle{vecArrow} = [thick, decoration={markings,mark=at position
    1 with {\arrow[semithick]{open triangle 60}}},
  double distance=1.4pt, shorten >= 5.5pt,
  preaction = {decorate},
  postaction = {draw,line width=1.4pt, white,shorten >= 4.5pt}]
\tikzstyle{innerWhite} = [semithick, white,line width=1.4pt, shorten >= 4.5pt]
\newcommand{\hdgstar}[2]{\node[star,star points=7,star point ratio=0.6,draw=black,fill=black!30] at (#1) {\huge\textbf{#2}};}
\newcommand{\hdgbottom}{0}
  
\begin{document}
  
  \title{Finite-Element Simulation of an Accelerator Magnet: an Exercise}
  \author{H. De~Gersem, I. Kulchytska-Ruchka and S. Sch\"ops\thanks
    {The exercise partially originate in the collaboration project \emph{Simulation of Transient Effects in Accelerator Magnets} (STEAM, https://espace.cern.ch/steam) between CERN and TEMF. This exercise has been carried out during lecture series at the RWTH University Aachen, at the TU Darmstadt and at the CERN Accelerator School. This work has been supported by the Excellence Initiative of the German Federal and State Governments and the Graduate School of Computational Engineering at TU Darmstadt.}}
  \institute{Institute for Accelerator Science and Electromagnetic Fields (TEMF), TU Darmstadt, Germany}
  
  \begin{abstract}
This note describes an extended exercise on the finite-element (FE) simulation of an accelerator magnet. The students construct and simulate a magnet model using the \texttt{FEMM} freeware. They get the opportunity to exercise on the theory of FEs, including Maxwell equations, magnetoquasistatic formulation, weighted residual approach, choice of appropriate FE shape functions and algebraic system of equations, thereby guided by fill-in sheets. They are invited to implement the most crucial parts of a simple FE solver. Finally, the own software is used to simulate the magnet once more and to develop some problem-dedicated post-processing routines. The exercise educates students in accelerator physics and electrical engineering on the construction and simulation of accurate and manageable FE models, the algorithms behind a standard FE solver and some ideas to extend a FE solver for own purposes. All necessary files to carry out the exercise are freely available.
  \end{abstract}
  
  
  \maketitle
    \thispagestyle{ARTTITLE}


\section{Introduction}

Finite-element (FE) simulation is indispensable in design procedures for normal-conducting and superconducting magnets. Many tools are available, as a commercial product or as freeware. The use of FE simulation, however, requires specific skills, which need to be educated consciously. Within that context, it makes sense that students get an insight in the interior of a FE solver. Moreover, customizing or even extending a FE solver for specific purposes can boost technological progress, e.g., in the field of accelerator magnets. This note describes an exercise aiming at these goals.

The note is structured as follows. In Section~II, the structure and content of the exercise is described. Section~III gives an overview of the provided material and hints for its usage. Section~IV reports about experience gained by tutors during the past years while guiding BSc- and MSc-students through the exercise. Section~V provides a summary and conclusion.

\section{Structure of the exercise}

The exercise is divided in 5~parts, which are preferably carried out at one exercise day, and in the prescribed order, but which also can be carried out individually and in different sessions. The exercise is accompanied by a self-explanatory guide (see Figs.~\ref{fig:sh1_title}--\ref{fig:sh28_newton2}). An introduction or explanation by a tutor before the start is not foreseen. It is assumed that the students attended the preceding lecture \cite{De-Gersem_2019ab}. The~worksheets try to address the students, invite them to carry out the tasks (numbered from \protect\tikz[baseline=-0.5ex]{
  \protect\node[star,star points=7,star point ratio=0.6,draw=black,fill=black!30] at (0,0) {1};} up to
\protect\tikz[baseline=-0.5ex]{
  \protect\node[star,star points=7,star point ratio=0.6,draw=black,fill=black!30] at (0,0) {41};}),
give hints which strategy is the most appropriate one and indicate how to interprete the results.

The first 4~worksheets (Figs.~\ref{fig:sh3_introduction}--\ref{fig:sh6_ourgoal}) describe the superconducting dipole magnet intended for bending the heavy-ion beam in the SIS-100 synchrotron of the Facility for Antiproton and Ion Research (FAIR, \texttt{www.fair.de} \cite{FAIR_2016aa}), which nears completion at the Helmholtzzentrum für Schwerionenforschung (GSI, Facility for Heavy Ion Research, \texttt{www.gsi.de} \cite{GSI_2016aa}) in Darmstadt, Germany.

\subsection{Part~1: Model construction}

The worksheets ''Geometry'', ''FEMM'', ''Materials'' and ''Boundary Conditions'' (Figs.~\ref{fig:sh7_geometry}--\ref{fig:sh10_boundaryconditions}) describe how to construct the magnet model from scratch using the graphical user interface of \texttt{FEMM} \cite{Meeker_2009aa}. For some students, this may be the first time confronted with a FE modeller. The model construction typically takes half an hour. When limited in time, one can decide to skip Part~1 and directly turn to Part~2. For that option, an already completed model is provided as well.

\subsection{Part~2: FE solution}

The worksheet with title ''Solve'' (Fig.~\ref{fig:sh11_solve}) guides the students through the solution process, thereby giving a bit of background information. The students are recommended to spend some time to the interpretation of the obtained magnetic flux distribution by a set of questions at the bottom of the worksheet. Here, the involvement of a tutor is known to improve the understanding. He/she should invite the students to play around with the software, thereby deliberately changing some parameters of the FE model and checking their impact on the solution. This part of the exercise takes only a quarter of an hour.

\subsection{Part~3: Mathematical derivation}

The worksheets ''Own implementation'' and ''Routines'' (Figs.~\ref{fig:sh12_ownimplementation} and~\ref{fig:sh13_routines}) already show a road map for the~implementation of an own FE solver. The students should have a look in the \texttt{FEMM} data files where they will recognize the data they have inserted for constructing the model and the mesh generated by the~built-in \texttt{Triangle} mesher \cite{Shewchuk_1996aa}. Thereafter, they get an overview about the provided \texttt{Matlab}/\texttt{Octave} \cite{Mathworks_2009aa,Eaton_2017aa} routines with which they will build an own FE solver. At this point, most probably, the student will have no idea how to start the implementation task. For that reason, the following worksheets offer the students the possibility to repeat the theory about magnetoquasistatic FE simulation, given during a~preceding lecture \cite{De-Gersem_2019ab}, on their own.

The worksheet ''Formulation'' (Fig.~\ref{fig:sh14_formulation}) guides the students from the Maxwell equations to the~magnetoquasistatic formulation as a function of the magnetic vector potential. In the worksheets ''Discretization'' (Figs.~\ref{fig:sh15_discretisation1}--\ref{fig:sh16_discretisation2}), the weighted residual approach and the FE shape functions are introduced. The~worksheet entitled ''Reduction to 2D'' (Fig.~\ref{fig:sh17_reduction}) instantiates the discrete formulation for the 2D cartesian case. The worksheets ''Compute Coefficients'' (Figs.~\ref{fig:sh18_computecoefficients1}--\ref{fig:sh19_computecoefficients2}) allow the students to calculate the~entries of the $3$-by-$3$ elementary matrices and $3$-by-$1$ elementary vector which after assembly form the FE system of equations. These derivations form the tasks~\protect\tikz[baseline=-0.5ex]{
  \protect\node[star,star points=7,star point ratio=0.6,draw=black,fill=black!30] at (0,0) {10};} to
\protect\tikz[baseline=-0.5ex]{
  \protect\node[star,star points=7,star point ratio=0.6,draw=black,fill=black!30] at (0,0) {27};}.
While this part of the~exercise is traditionally experienced as difficult, this part has been divided in a large number of small tasks. This part takes $0.5$ to $1.5$ hours of time.

\subsection{Part~4: Implementation}

Task~\protect\tikz[baseline=-0.5ex]{
  \protect\node[star,star points=7,star point ratio=0.6,draw=black,fill=black!30] at (0,0) {28};}
asks for a first implementation step. In the routines \texttt{curlcurl\_ll.m}, \texttt{edgemass\_ll.m} and \texttt{current\_Pstr.m}, the students find places marked by the comment ''IMPLEMENTATION POINT'' where they should code the matrix entries as calculated before. This task is crucial in the whole exercise. The tutor should make sure that all students reach this task and accomplish it successfully. The~worksheets ''Magnetic Flux Density'' (Fig.~\ref{fig:sh20_magnfluxdens}) and ''Magnetic Energy'' (Fig.~\ref{fig:sh21_magnenergy}) require similar derivations and implementations. Up to this point, no own FE solution is required. The implemented routines are tested by post-processing the magnetic energy and the~magnetic flux density from the solution for the~magnetic vector potential obtained by \texttt{FEMM}.

The worksheet ''Boundary Conditions'' (Fig.~\ref{fig:sh22_boundaryconditions}) gives basic information about inserting Dirichlet boundary conditions. The treatment of the boundary conditions is given as a ready implementation (routines \texttt{shrink.m} and \texttt{inflate.m}). After inserting the boundary conditions, the system of equations can be solved and a solution close to the one of \texttt{FEMM} should come out. For the complete part~4, students typically need $1$ to $2$ hours. 

\subsection{Part~5: Post-processing}

Part~5 of the exercise comprises a post-processing step dedicated to accelerator-magnet design, i.e., the~calculation of the \textit{harmonic components} and the \textit{skew harmonic components} of the magnet's aperture field. The theory is written down on the worksheet ''Aperture Field (1/2)'' (Fig.~\ref{fig:sh23_aperturefield1}), whereas the guidance for implementation is organized as three tasks on the next worksheet (Fig.~\ref{fig:sh24_aperturefield2}).

The last part of the exercise extends the own FE solver to deal with nonlinear magnetic materials. Possible linearization techniques are described in the worksheet ''Nonlinear Material Properties'' (Fig.~\ref{fig:sh25_nonlinearmaterialproperties}). The successive-substitution and Newton algorithms are explained in the three last worksheets (Figs.~\ref{fig:sh26_successivesubstitution}--\ref{fig:sh28_newton2}), together with indication for implementation of the algorithm in the main file of the own software. The outcome is a renewed calculation of the (skew) harmonic components of the aperture field, now accounting for the inevitable saturation of the magnet yoke. The results gathered in the table of Fig.~\ref{fig:sh28_newton2} should show a decrease of the magnitudes of the higher-harmonic components compared to the results collected in the table of Fig.~\ref{fig:sh24_aperturefield2}. This illustrates that the SIS-100 magnet has been optimized such that the aperture field is maximally homogeneous for the maximal field \cite{Kovalenko_2002aa}.

Part~5 takes a total of $1$ up to $2$ hours for completion.

\section{Exercise material}

The material needed to carry out the exercise can be found in \cite{De-Gersem_2019ad}. The worksheets supporting the exercise are also contained within this note, see Figs.~\ref{fig:sh1_title}--\ref{fig:sh28_newton2}. The figure captions give a few didactical hints, which come from the tutors who guided the exercise during the past years. Together with that, the students can download the software. The material is divided in 5~independent parts, which allows the students to enter the exercise at any point, e.g., skipping the construction of the model (part~1) and directly start calculating (part~2), or skipping the mathematical part (part~3) and directly passing from the calculation using \texttt{FEMM} (part~2) to the implementation of an own FE solver (part~4). The material also included the~slides of the preceding lecture.

The version of the software provided for part~4 of the exercise is deliberately made incomplete. At clearly marked so-called ''IMPLEMENTATION POINTS'', the students are invited to complete the~code. The code fragments are chosen in order to support insight in the key parts of a FE solver, thereby avoiding all parts which are solely intended for \textit{bookkeeping} (handling mesh data, boundary conditions, material distributions, excitations). The software provided for part~5 is, however, complete. Uninterested student could decide to copy-paste the corresponding code fragments -- from which they would not learn that much. The completeness of the provided material, however, makes sure that all students return home with a fully functional 2D nonlinear magnetostatic FE solver.

\pagebreak
\section{Discussion}

The exercise has been used as part of several lecture series at several universities. By that, experience has been gathered, which is shared in this section.

Upon invitation to work in groups, the students typically form groups of two or three, but do not divide the work. Instead, they work on one computer through the worksheets from start to finish. This is beneficial for guard against errors and improving understanding, but requires maximal working time, amounting to $4$ hours for a team of skilled and motivated students up to $12$ hours for unexperienced BSc students. When time is limited, the tutors should actively suggest working in parallel.

When the students create the FE model themselves, a lot of modelling errors occur. The most common ones are wrong units, wrong assignment of material properties, wrong assignment of exciting currents and wrong assignment of boundary conditions (of course, this would be slightly different when another software would be used). Often, the students need a little help to find the error. Mostly, it is sufficient to force them to look at the (wrong) magnetic flux distribution, point them explicitly to the erroneous region and ask them what could be the origin thereof. If the students are proficient in FE modelling, the first part of the exercise can be skipped.

The students hesitate to start the implementation part (part~3), which is particularly caused by the size of the FE software (some dozens of \texttt{Matlab}/\texttt{Octave} routines) and the lack of programming experience. Here, the tutor should recommend to start with the implementation tasks one by one and in the given order. The implementation points in the software are well marked and stick closely to the~mathematical derivation. This should keep the hurdle as low as possible. A related observation is made when the students finish the implementation part. Then, they are surprised to be successful and get more confident in FE modelling and simulation. Some students ask for hints for further implementation of the more advanced methods discussed in the lecture. Other students try to adapt the software for own purposes, e.g., for solving an example related to their BSc- or MSc-thesis work.

The implementation of the post-processing routine for calculating the harmonic and skew-harmonic components of the aperture field is tough. Students with a pronounced interest in accelerator physics remain motivated. However, for students with focus on studying magnetic field simulation by FE methods, this parts feels superfluous. The tutor may invite the latter to proceed to the treatment of the nonlinear case and use the magnetic energy as a reference quantity of interest instead of the harmonic and skew-harmonic components.

\section{Summary and conclusions}

This paper describes an extended exercise where students in physics, mathematics and engineering are invited to set up a 2D finite-element (FE) model of a superconducting accelerator magnet. Furthermore, the exercise briefly recalls standard FE theory and enables the students to bring this knowledge into praxis by implementing a few parts in a given FE software framework. The exercise ends by adding a~post-processing tool to derive the harmonic and skew-harmonic field components from the FE solution for a superconducting magnet model. The exercise takes $4$ hours for skilled and motivated MSc-students up to $12$ for unexperienced BSc-students. The exercise can be carried out in parts. The main learning achievement is the fact that students get convinced about the capabilities of FE simulation and get confident for future tasks involving FE simulation.

\section*{Acknowledgements}

We thank Wolfgang Ackermann, Markus Borkowski, Thorben Casper, Idoia Cortes Garcia, Laura A.M. D'Angelo, Erion Gjonaj, Dimitrios Loukrezis, Nicolas Marsic, Andreas Pels and Erik Schnaubelt for their contributions to this field of research and for the development and maintenance of the software. Moreover, we thank Gregor Bavendiek, Christian Krüttgen, Fabian Müller, Peter Offermann, Aryanti Putri and Dries Vanoost for their dedication as tutors in the exercise classes. We thank all students of TU~Darmstadt, RWTH Aachen, KU~Leuven, Université de Lille and all participants of the CERN Accelerator School for their suggestions for improvement of the worksheets. Last but not least, we thank David Meeker and Jonathan Shewchuk for putting \texttt{FEMM} \cite{Meeker_2009aa} and \texttt{Triangle} \cite{Shewchuk_1996aa} as freeware on the world wide web.

\begin{figure}[t]
  \centering
  \begin{tikzpicture}

\draw[rounded corners] (0,\hdgbottom) rectangle (16,20);

\node [anchor=north west] at (0,20) {\begin{minipage}{16cm}\centering%
  \vspace{4cm}{\Huge Magnetostatic Simulation \\ of an Accelerator Magnet}\\
  {\Large\vspace{2cm} make your own finite-element solver\\
    \vspace{2cm} CERN Accelerator School\\
    11-23 November 2018\\
    Thessaloniki, Greece\\
    \vspace{2cm} Prof. Dr.-Ing. Herbert De Gersem \\
    Iryna Kulchytska-Ruchka, M.Sc.\\
    Prof. Dr. rer. nat. Sebastian Schöps\\
    \vspace{1cm} Institute for Accelerator Science and Electromagnetic Fields (TEMF)\\
    TU Darmstadt \\ Germany}
  \end{minipage}};

\end{tikzpicture}%
  \caption{Title sheet.}
  \label{fig:sh1_title}
\end{figure}
\begin{figure}[t]
  \centering
  \begin{tikzpicture}
\Large
\draw[rounded corners] (0,\hdgbottom) rectangle (16,20);
\node[anchor=west, fancytitle, rounded corners] at (2,20) {\textbf{Introduction}};

\node [anchor=north west] at (0.5,19) {\begin{minipage}{15cm}\begin{flushleft}%
  \vspace{0.0cm}what we will do ...
  \begin{itemize}[nolistsep]
    \vspace{0.2cm}\item[] In this exercise, we start from the Maxwell equations and finish at the magnetic field in an accelerator magnet. Except for the construction of the geometry and the definition of the materials, we will implement all steps ourselves.
  \end{itemize}
  \vspace{0.5cm}make sure that you bring with you ...
  \begin{enumerate}[nolistsep]
    \vspace{0.2cm}\item your laptop,
    \item a recent version of FEMM (\texttt{www.femm.info/}),
    \item Matlab (\texttt{www.mathworks.com}) or Octave (\texttt{http://www.gnu.org/software/octave/})
    \item the slides of the previous lectures.
  \end{enumerate}
  \vspace{0.5cm}work together ...
  \begin{enumerate}[nolistsep]
    \vspace{0.2cm}\item in small groups (2 or 3 persons).
    \item divide the work,
    \item but keep informed about the “what and how” the other is doing.
    \item adapt a careful attitude, double check everything, in numerical simulation, a small error is sufficient to blow up the universe.
  \end{enumerate}
  \vspace{0.5cm}the goal of an exercise is to learn, therefore ...
  \begin{enumerate}[nolistsep]
    \vspace{0.2cm}\item ask everything you do not know or you are not sure of.
    \item repeat an exercise (maybe with slightly different parameters) until you understand its solution thoroughly.
  \end{enumerate}
  \vspace{0.5cm}enjoy ...
  \end{flushleft}\end{minipage}};

\end{tikzpicture}%
  \caption{Introductory sheet: summary, software installation, goal, strategy. This sheet should attract the students' attention and motivate them to start with the exercise.}
  \label{fig:sh3_introduction}
\end{figure}
\begin{figure}[t]
  \centering
  \input{degersem2_sheet3}
  \caption{Context: dipole accelerator magnets. For most students, this worksheet does not bring new information. It recapitalizes knowledge from a first-semester physics course. The worksheet makes a direct link between theory and the particular application.}
  \label{fig:sh4_acceleratormagnet}
\end{figure}
\begin{figure}[t]
  \centering
  \input{degersem2_sheet4}
  \caption{Background of the SIS-100 superconducting dipole magnet. This background knowledge and the provided links invites the interested students to have a look to the larger context.}
  \label{fig:sh5_background}
\end{figure}
\begin{figure}[t]
  \centering
  \input{degersem2_sheet5}
  \caption{Goals of the exercise and strategy to accomplish these goals. This worksheet provides a task description. The goals range beyond the purpose of this exercise (which are getting familiar with the FE method for magnetic field simulation) but trigger students with a clear interest in accelerator physics.}
  \label{fig:sh6_ourgoal}
\end{figure}
\begin{figure}[t]
  \centering
  \input{degersem2_sheet6}
  \caption{Geometry and dimensions of the SIS-100 superconducting dipole magnet. This worksheet enables the~students to build up the model themselves. Nonetheless, a ready model can be offered to bring down the extend of the~exercise or to serve as a backup. The steps to be carried out by the students are marked by \protect\tikz[baseline=-0.5ex]{
    \protect\node[star,star points=7,star point ratio=0.6,draw=black,fill=black!30] at (0,0) {1};
  } up to
  \protect\tikz[baseline=-0.5ex]{
    \protect\node[star,star points=7,star point ratio=0.6,draw=black,fill=black!30] at (0,0) {41};
  }. This allows the tutor to monitor the progress of the students or the student groups.
}
  \label{fig:sh7_geometry}
\end{figure}
\begin{figure}[t]
  \centering
  \input{degersem2_sheet7}
  \caption{Overview of the most frequently used parts of \texttt{FEMM}. For further information, the students are referred to the user manual or the tutorials.}
  \label{fig:sh8_femm}
\end{figure}
\begin{figure}[t]
  \centering
  \input{degersem2_sheet8}
  \caption{Material: nonlinear iron for the magnet's yoke. It is recommended to carry out a first simulation with a~linear material and a second one with the true nonlinear material. One should find that the field quality is better for the true nonlinear case than for the linear case.}
  \label{fig:sh9_materials}
\end{figure}
\begin{figure}[t]
  \centering
  \input{degersem2_sheet9}
  \caption{Boundary conditions: explanation about the different types of boundary conditions, implementation of the boundary conditions. Boundary conditions typically raise many questions. The tutor may invite the student to try different boundary conditions and interprete the resulting magnetic flux distributions.}
  \label{fig:sh10_boundaryconditions}
\end{figure}
\begin{figure}[t]
  \centering
  \input{degersem2_sheet10}
  \caption{Procedure for simulation the magnet FE model. A bit of background information is given about the mesher and the solver. It is important that the students are persuaded to carefully look at the resulting magnetic flux distributions. The tutor should point to the magnetic flux lines refracting at material interfaces, to the magnetic flux lines encircling the wires, to the magnetic flux lines touching the~boundary normally or tangentially and to regions with high saturation.}
  \label{fig:sh11_solve}
\end{figure}
\begin{figure}[t]
  \centering
  \begin{tikzpicture}
\Large
\draw[rounded corners] (0,\hdgbottom) rectangle (16,20);
\node[anchor=west, fancytitle, rounded corners] at (2,20) {\textbf{Own Implementation}};
\node[anchor=north east] at (16,20) [] {\begin{minipage}{15cm}\begin{flushright} doing it yourself ...\end{flushright}\end{minipage}};

\node [anchor=north west] at (0.5,19) {\begin{minipage}{12.5cm}\begin{flushleft}%
\begin{itemize}
\item formulation
\item discretisation
\item implementation in Matlab
\begin{enumerate}
\item 2D, linear, magnetostatic solver
\begin{itemize}[nolistsep]
\item getting information out of FEMM data files
\item check this information (e.g. using figures)
\item constructing the system of equations
\item applying boundary conditions
\item solving the system
\item post-processing
\item write the solution to a FEMM data file
\end{itemize}
\item 2D, nonlinear, magnetostatic solver
\begin{itemize}[nolistsep]
\item reading a nonlinear material characteristic
\item finding elements belonging to the iron parts
\item setting up a nonlinear iteration by successive substitution and/or Newton
\end{itemize}
\end{enumerate}
\end{itemize}
\end{flushleft}\end{minipage}};

\draw[thick, decoration={markings,mark=at position
  1 with {\arrow[semithick]{triangle 60}}},
double distance=2.4pt, shorten >= 5.5pt,
preaction = {decorate},
postaction = {draw,line width=1.4pt, blue,shorten >= 4.5pt}] (1,19) to (1,7);
\node [rotate=90] at (0.5,13) {road map};

\node [anchor=north west] at (2,4) {\begin{minipage}{11.5cm}\begin{flushleft}%
  \begin{itemize}[nolistsep]
  \item take a look to the \texttt{FEMM} data files (extensions \texttt{.fem} and \texttt{.ans}). Figure out which information is needed for setting up an own finite-element solver.
  \end{itemize}
  \end{flushleft}\end{minipage}};
\hdgstar{1.5,3}{8}

\node [anchor=north west] at (2,1.5) {\begin{minipage}{11.5cm}\begin{flushleft}%
  \begin{itemize}[nolistsep]
  \item some of the auxiliary routines are provided.
  \item write the main routines yourself!
  \end{itemize}
  \end{flushleft}\end{minipage}};
\hdgstar{1.5,0.8}{9}

\end{tikzpicture}%
  \caption{Road map to come to an own implementation of a FE solver. This worksheet points to the third phase of the exercise, i.e., writing an own FE solver for carrying out the same simulation starts. The students are invited to have a look in the \texttt{FEMM} data file, in which they should find all data related to the FE model, e.g., the mesh. The~implementation makes use of auxiliary routines, which are provided.}
  \label{fig:sh12_ownimplementation}
\end{figure}
\begin{figure}[t]
  \centering
  \begin{tikzpicture}
\large
\draw[rounded corners] (0,\hdgbottom) rectangle (16,20);
\node[anchor=west, fancytitle, rounded corners] at (2,20) {\textbf{Routines}};

\node [anchor=north west] at (0.5,19) {\begin{minipage}{15cm}\begin{flushleft}%
\begin{tabular}{|p{6.5cm}|p{7.7cm}|}
\hline
\uline{\texttt{curl.m}}                     & \uline{computing $\vec{B}$ from $\vec{A}$} \\
\uline{\texttt{curlcurl\_ll.m}}             & \uline{assembling a curl-curl stiffness matrix for linear reluctivities} \\
\uline{\texttt{curlcurl\_ll\_nonlinear.m}}  & \uline{assembling a curl-curl stiffness matrix and magnetisation vector for nonlinear reluctivities} \\
\uline{\texttt{current\_Pstr.m}}            & \uline{assembling right-hand-side vectors for coils} \\
\uline{\texttt{driver.m}}                   & \uline{main file} \\
\uline{\texttt{edgemass\_ll.m}}             & \uline{assembling a mass matrix for linear conductivities} \\
\texttt{findlab.m}                          & find the indices in an array of strings where a certain label is found \\
\texttt{mesh\_linear\_shape \_functions.m}  & compute all information for linear finite-element shape functions \\
\texttt{nlin\_evaluate.m}                   & evaluate a nonlinear material characteristic using a data structure for nonlinear materials \\
\texttt{nlin\_initialise.m}                 & initialise a data structure for nonlinear materials \\
\texttt{plot\_point.m}                      & plot geometry points \\
\texttt{ppder.m}                            & derive a spline-polynomial \\
\texttt{pyth.m}                             & compute the magnitude of a vector field \\
\texttt{read\_femm.m}                       & read a problem from a \texttt{femm} \texttt{.ans} file \\
\texttt{save\_femm.m}                       & save a problem to a \texttt{femm} \texttt{.ans} file \\
\texttt{savedivide.m}                       & divide by vectors and deal with zero denominators \\
\texttt{sis100\_geometry.m}                 & drivers from creating the SIS100 geometry \\
\texttt{viewequi.m}                         & draw an equipotential plot \\
\texttt{write\_femm\_geometry.m}            & write geometric information to a \texttt{femm} \texttt{.fem} file \\
\hline
\end{tabular}

\vspace{0.5cm}\begin{itemize}[nolistsep]
\item more detailed information can be obtained by typing ''help filename'' in \texttt{Matlab}.
\item files in which you should add some implementation are underlined. You have to insert some implementation at places indicated by ''IMPLEMENTATION POINT''.
\end{itemize}
  \end{flushleft}\end{minipage}};

\end{tikzpicture}%
  \caption{Overview of the provided \texttt{Matlab} routines. Most of the routines are given in full. In the underlined routines, one or several \emph{implementation points} are inserted, at which the students should add own coding. Complete routines are provided as well, such that the students can check their implementation or skip the implementation step. At this point, most students feel overburdened. For that reason, in an intermediate step starting from the~following worksheet, they are invited to rehearse the theory.}
  \label{fig:sh13_routines}
\end{figure}
\begin{figure}[t]
  \centering
  \begin{tikzpicture}
\Large
\draw[rounded corners] (0,\hdgbottom) rectangle (16,20);
\node[anchor=west, fancytitle, rounded corners] at (2,20) {\textbf{Formulation}};
\node[anchor=north east] at (16,20) [] {\begin{minipage}{10cm}\begin{flushright} where we turn the Maxwell equations into the magnetoquasistatic formulation as a function of the magnetic vector potential ...\end{flushright}\end{minipage}};

\draw [mybox] (0.5,3) rectangle (15.5,0.5);
\draw[vecArrow] (14,15) to (14,3);

\draw [mybox] (0.5,19) rectangle (5.5,9);
\node[fancytitle, rounded corners] at (3,19) {Maxwell's equations};
\node[anchor=west] at (0.5,18) [] (box1) {\begin{minipage}{5cm}\begin{flalign*}
  \nabla\cdot\vec{D}  &= \rho &
  \end{flalign*}\end{minipage}};
\node[anchor=west] at (0.5,15.5) [] (box2) {\begin{minipage}{5cm}\begin{flalign*}
  \nabla\times\vec{H} &= \vec{J}+\frac{\partial\vec{D}}{\partial t} &
  \end{flalign*}\end{minipage}};
\node[anchor=west] at (0.5,13) [] (box3) {\begin{minipage}{5cm}\begin{flalign*}
  \nabla\cdot\vec{B}  &= 0 &
  \end{flalign*}\end{minipage}};
\node[anchor=west] at (0.5,10.5) [] (box4) {\begin{minipage}{5cm}\begin{flalign*}
  \nabla\times\vec{E} &= -\frac{\partial\vec{B}}{\partial t} &
  \end{flalign*}\end{minipage}};

\draw [mybox] (0.5,8.5) rectangle (5.5,3.5);
\node[fancytitle, rounded corners] at (3,8.5) {material laws};
\node[anchor=west] at (0.5,7.5) [] (box5) {\begin{minipage}{5cm}\begin{flalign*}
  \vec{D} &= \varepsilon\vec{E} &
  \end{flalign*}\end{minipage}};
\node[anchor=west] at (0.5,6) [] (box6) {\begin{minipage}{5cm}\begin{flalign*}
  \vec{B} &= \mu\vec{H} =\frac{1}{\nu}\vec{H} &
  \end{flalign*}\end{minipage}};
\node[anchor=west] at (0.5,4.5) [] (box7) {\begin{minipage}{5cm}\begin{flalign*}
  \vec{J} &= \sigma\vec{E} &
  \end{flalign*}\end{minipage}};
\node[fancytitle, rounded corners] at (5.5,6) {$4$};
\node[fancytitle, rounded corners] at (5.5,4.5) {$5$};

\draw [mybox] (10,17) rectangle (15.5,15);
\draw[vecArrow] (5.5,15.2) to (10,15.2);
\node[anchor=south west] at (5.5,15.2) [] {\begin{minipage}{4.0cm} \normalsize neglect electrical energy w.r.t. magnetic energy and losses \end{minipage}};
\node[fancytitle, rounded corners] at (15.5,15) {$1$};
\hdgstar{15,17}{10}

\draw [mybox] (10,14) rectangle (15.5,12);
\draw[vecArrow] (5.5,12.7) to (10,12.7);
\node[anchor=south west] at (5.5,12.7) [] {\begin{minipage}{4.0cm} \normalsize integrate in space and define magnetic vector potential \end{minipage}};
\node[fancytitle, rounded corners] at (15.5,12) {$2$};
\hdgstar{15,14}{11}

\draw [mybox] (10,11) rectangle (15.5,9);
\draw[vecArrow] (5.5,10.2) to (10,10.2);
\node[anchor=south west] at (5.5,10.2) [] {\begin{minipage}{4.0cm} \normalsize integrate in space and define electric scalar potential\end{minipage}};
\node[fancytitle, rounded corners] at (15.5,9) {$3$};
\hdgstar{15,11}{12}

\draw[vecArrow] (5.75,6) to (14,6);
\draw[vecArrow] (5.75,4.5) to (14,4.5);
\node[anchor=south west] at (8,6) [] {\begin{minipage}{5.0cm} combine \tikz{\node[fancytitle, rounded corners] {$1$};}-\tikz{\node[fancytitle, rounded corners] {$5$};}\end{minipage}};
\node[fancytitle, rounded corners, anchor=west] at (6,3) {magnetoquasistatic formulation};
\hdgstar{15,3}{13}

\end{tikzpicture}%
  \caption{Worksheet for deriving the magnetoquasistatic formulation from the Maxwell equations. The worksheet allows to repeat theory step by step. This is preferably carried out using the suggestions provided on the worksheet but can also be accomplished by copying from the slides shown during the lecture.}
  \label{fig:sh14_formulation}
\end{figure}
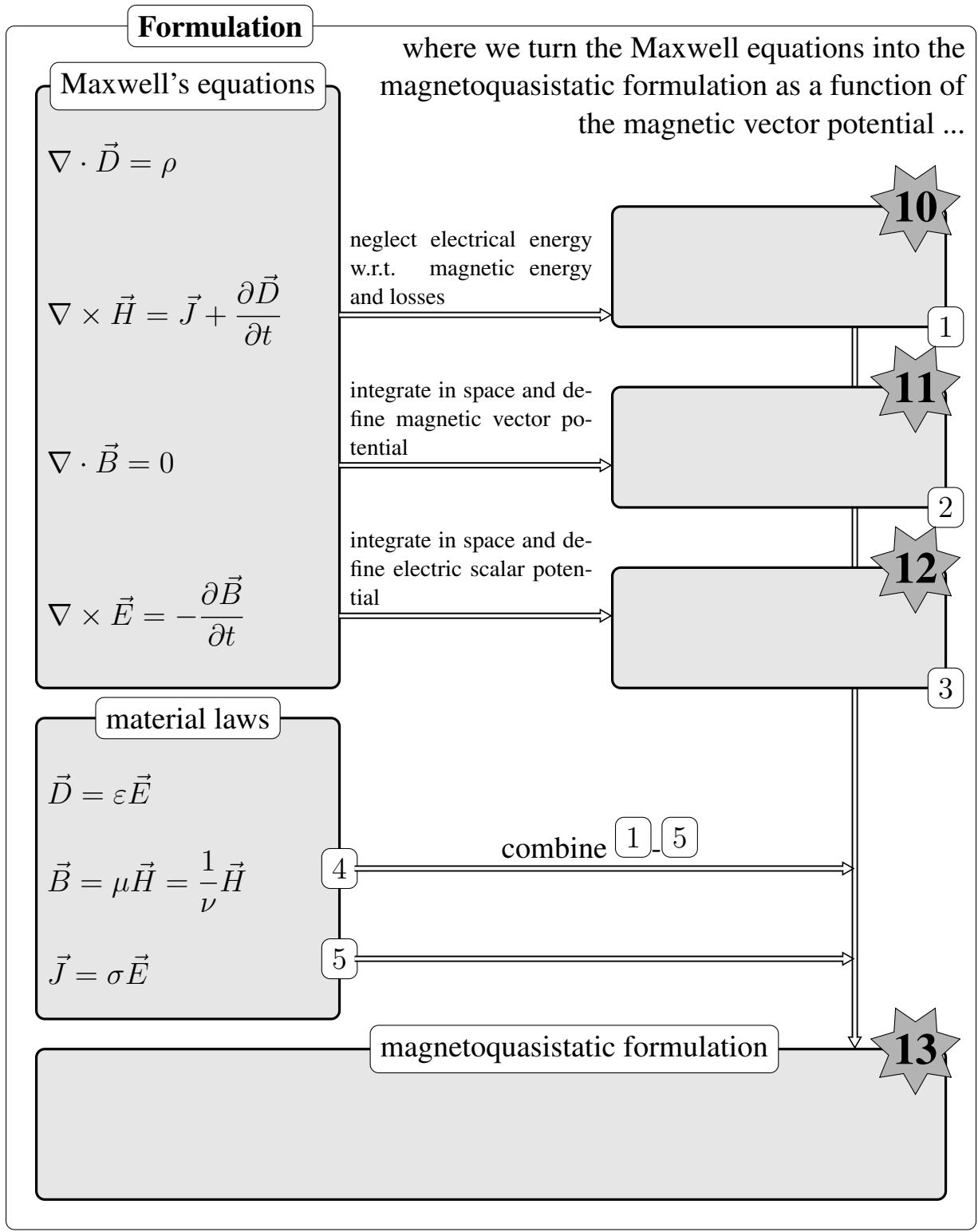
\begin{figure}[t]
  \centering
  \begin{tikzpicture}
\Large
\draw[rounded corners] (0,\hdgbottom) rectangle (16,20);
\node[anchor=west, fancytitle, rounded corners] at (2,20) {\textbf{Discretisation (1/2)}};
\node[anchor=north east] at (16,20) [] {\begin{minipage}{15cm}\begin{flushright} from the continuous to the discrete level ...\end{flushright}\end{minipage}};

\draw [mybox] (0.5,19) rectangle (15.5,16.5);
\node [anchor=north west] at (0.5,19) {\begin{minipage}{13.5cm}\begin{flushleft} (strong) formulation\\
  \vspace{-0.5cm}\begin{flalign*}
    \nabla\times\left(\nu\nabla\times\vec{A}\right)+\sigma\frac{\partial\vec{A}}{\partial t} &=\vec{J}_\text{s} &
  \end{flalign*}
  \end{flushleft}\end{minipage}};

\hdgstar{1,15.75}{14}
\draw[vecArrow] (2,16.5) to (2,15);
\draw [mybox] (0.5,15) rectangle (15.5,12.5);
\node [anchor=west] at (2.5,15.75) [] {\begin{minipage}{13.5cm}\begin{flushleft} apply the weighted residual method, i.e., (a) multiply with test functions $\vec{w}_i(\vec{r})$; (b) integrate over the computational domain
$V$. \end{flushleft}\end{minipage}};

\hdgstar{1,11.75}{15}
\draw[vecArrow] (2,12.5) to (2,11);
\draw [mybox] (0.5,11) rectangle (15.5,8.5);
\node [anchor=west] at (2.5,11.75) [] {\begin{minipage}{13.5cm}\begin{flushleft} apply the vector-calculus formula $(\nabla\times\vec{v})\cdot\vec{w} =\nabla\cdot(\vec{v}\times\vec{w})+\vec{v}\cdot(\nabla\times\vec{w})$. \end{flushleft}\end{minipage}};

\hdgstar{1,7.75}{16}
\draw[vecArrow] (2,8.5) to (2,7);
\draw [mybox] (0.5,7) rectangle (15.5,4.5);
\node [anchor=west] at (2.5,7.75) [] {\begin{minipage}{13.5cm}\begin{flushleft} apply Gauss theorem $\int_V\nabla\cdot\vec{v}\,\mathrm{d}V =\int_{\partial V}\vec{v}\cdot\,\mathrm{d}\vec{A}$. \end{flushleft}\end{minipage}};

\hdgstar{1,3.75}{17}
\draw[vecArrow] (2,4.5) to (2,3);
\draw [mybox] (0.5,3) rectangle (15.5,0.5);
\node [anchor=west] at (2.5,3.75) [] {\begin{minipage}{13.5cm}\begin{flushleft} assume that only (a) Dirichlet boundary conditions and \\ (b) homogeneous Neumann boundary conditions are applied. \end{flushleft}\end{minipage}};

\end{tikzpicture}%
  \caption{Worksheet for deriving the weak form of the magnetoquasistatic formulation. Again, the students can follow the provided suggestions or can copy the appropriate parts of the lecture.}
  \label{fig:sh15_discretisation1}
\end{figure}
\begin{figure}[t]
  \centering
  \begin{tikzpicture}
\Large
\draw[rounded corners] (0,\hdgbottom) rectangle (16,20);
\node[anchor=west, fancytitle, rounded corners] at (2,20) {\textbf{Discretisation (2/2)}};
\node[anchor=north east] at (16,20) [] {\begin{minipage}{15cm}\begin{flushright} and end up with a system of equations ...\end{flushright}\end{minipage}};

\draw [mybox] (0.5,19) rectangle (15.5,16.5);
\node [anchor=north west] at (0.5,19) {\begin{minipage}{13.5cm}\begin{flushleft} weak formulation\\
  \vspace{-0.5cm}\begin{flalign*}            \int_V\nu\nabla\times\textcolor{blue}{\vec{A}}\cdot\nabla\times\textcolor{red}{\vec{w}_i}\,\mathrm{d}V
  +\int_V\sigma\frac{\partial\textcolor{blue}{\vec{A}}}{\partial t}\cdot\textcolor{red}{\vec{w}_i}\,\mathrm{d}V
  &=\int_V\vec{J}_\text{s}\cdot\textcolor{red}{\vec{w}_i}\,\mathrm{d}V & \end{flalign*}\end{flushleft}\end{minipage}};

\hdgstar{1,15.75}{18}
\draw[vecArrow] (2,16.5) to (2,15);
\draw [mybox] (0.5,15) rectangle (15.5,12.5);
\node [anchor=west] at (2.5,15.75) [] {\begin{minipage}{13.5cm}\begin{flushleft} insert the discretisation for the magnetic vector potential $\textcolor{blue}{\vec{A}(\vec{r})}=\sum_j u_j\textcolor{blue}{\vec{w}_j(\vec{r})}$. \end{flushleft}\end{minipage}};

\hdgstar{1,7.75}{19}
\draw[vecArrow] (2,12.5) to (2,3);
\node [anchor=west] at (2.5,11.75) [] {\begin{minipage}{13.5cm}\begin{flushleft} find expressions for the coefficients. \end{flushleft}\end{minipage}};
\draw [mybox] (3.5,11)  rectangle (15.5,9);
\node [anchor=west] at (3.5,10.25) {\begin{minipage}{6cm}\begin{flushleft}\begin{flalign*} \mathbf{K}_{\nu,ij}&= & \end{flalign*}\end{flushleft}\end{minipage}};
\draw [mybox] (3.5,8.75) rectangle (15.5,6.75);
\node [anchor=west] at (3.5,8) {\begin{minipage}{6cm}\begin{flushleft}\begin{flalign*} \mathbf{M}_{\sigma,ij}&= & \end{flalign*}\end{flushleft}\end{minipage}};
\draw [mybox] (3.5,6.5) rectangle (15.5,4.5);
\node [anchor=west] at (3.5,5.75) {\begin{minipage}{6cm}\begin{flushleft}\begin{flalign*} \mathbf{f}_{\mathrm{s},i}&= & \end{flalign*}\end{flushleft}\end{minipage}};
\node [anchor=west] at (2.5,3.75) [] {\begin{minipage}{13.5cm}\begin{flushleft} write down the resulting algebraic system of equations. \end{flushleft}\end{minipage}};
\draw [mybox] (0.5,3) rectangle (15.5,0.5);

\end{tikzpicture}%
  \caption{Worksheet for deriving the algebraic system of equations after applying the FE method to the weak form of the magnetoquasistatic formulation. The tutor can indicate that the arising system of equations enables the~implemenation of the FE method on a computer, thereby making a link to the third part of the exercise.}
  \label{fig:sh16_discretisation2}
\end{figure}
\begin{figure}[t]
  \centering
  \begin{tikzpicture}
\Large
\draw[rounded corners] (0,\hdgbottom) rectangle (16,20);
\node[anchor=west, fancytitle, rounded corners] at (2,20) {\textbf{Reduction to 2D}};
\node[anchor=north east] at (16,20) [] {\begin{minipage}{15cm}\begin{flushright} simplify ...\end{flushright}\end{minipage}};

\node [anchor=north west] at (0,19) {\begin{minipage}{15.5cm}\begin{flushleft}%
  \begin{itemize}[nolistsep]
  \item when the geometry/excitation/boundary conditions remain the same along the axis of the device,
  \item when the current is perpendicular to the cross-section,
  \item then the flux is aligned with the cross-section,
  \item\vspace{1.1cm} then
  \item\vspace{1.9cm} and
  \item\vspace{1.0cm} apply the edge functions
  \begin{flalign*}\hspace{1cm}\textcolor{blue}{\vec{w}_j} &=\textcolor{blue}{\frac{N_j(x,y)}{\ell_z}\vec{e}_z} &
  \end{flalign*}
  both as test and trial functions,
  \item\vspace{1.2cm} then
  \item\vspace{1.9cm} and
  \item\vspace{1.9cm} and
  \end{itemize}
  \end{flushleft}\end{minipage}};

\draw [mybox] (2.5,16.0) rectangle (15.5,14.0);
\node [anchor=west] at (2.5,15.25) {\begin{minipage}{6cm}\begin{flalign*} \vec{A}(x,y,z)&=(\quad\quad,\quad\quad,\quad\quad)& \end{flalign*}\end{minipage}};
\hdgstar{15.5,15.0}{20}

\draw [mybox] (2.5,13.5) rectangle (15.5,11.5);
\node [anchor=west] at (2.5,12.75) {\begin{minipage}{6cm}\begin{flalign*}\vec{B}(x,y,z)&=(\quad\quad,\quad\quad,\quad\quad)& \end{flalign*}\end{minipage}};
\hdgstar{15.5,12.5}{21}

\draw [mybox] (2.5,7.5) rectangle (15.5,5.5);
\node [anchor=west] at (2.5,6.75) {\begin{minipage}{6cm}\begin{flalign*} \mathbf{K}_{\nu,ij}&= & \end{flalign*}\end{minipage}};
\hdgstar{15.5,6.5}{22}

\draw [mybox] (2.5,5.0) rectangle (15.5,3.0);
\node [anchor=west] at (2.5,4.25) {\begin{minipage}{6cm}\begin{flalign*} \mathbf{M}_{\sigma,ij}&= & \end{flalign*}\end{minipage}};
\hdgstar{15.5,4.0}{23}

\draw [mybox] (2.5,2.5) rectangle (15.5,0.5);
\node [anchor=west] at (2.5,1.75) {\begin{minipage}{6cm}\begin{flalign*} \mathbf{f}_{\mathrm{s},i}&= & \end{flalign*}\end{minipage}};
\hdgstar{15.5,1.5}{24}

\end{tikzpicture}%
  \caption{Worksheet for adapting the formulation to the 2D case. The instantiation of the general 3D FE method to the cartesian 2D case should make the students more familiar with the FE method.}
  \label{fig:sh17_reduction}
\end{figure}
\begin{figure}[t]
  \centering
  \input{degersem2_sheet17}
  \caption{Worksheet for explaining the computation of coefficients for hat FE functions. The general integration rule for lowest-order 2D hat functions allows to calculate all elementary matrices asked for in the next worksheets.}
  \label{fig:sh18_computecoefficients1}
\end{figure}
\begin{figure}[t]
  \centering
  \begin{tikzpicture}
\Large
\draw[rounded corners] (0,\hdgbottom) rectangle (16,20);
\node[anchor=west, fancytitle, rounded corners] at (2,20) {\textbf{Compute Coefficients (2/2)}};
\node[anchor=north east] at (16,20) [] {\begin{minipage}{15cm}\begin{flushright} further down to the implementation ...\end{flushright}\end{minipage}};

\node [anchor=north west] at (0,19) {\begin{minipage}{15cm}\begin{flushleft}%
  Compute the local coefficient matrices and vectors:
  \end{flushleft}\end{minipage}};

\draw [mybox] (0.5,18) rectangle (15.5,16);
\node [anchor=west] at (0.5,17.25) {\begin{minipage}{6cm}\begin{flalign*} \mathbf{K}_{\nu,ij}^{(e)}&= & \end{flalign*}\end{minipage}};
\hdgstar{15.5,17}{25}

\draw [mybox] (0.5,15) rectangle (15.5,13);
\node [anchor=west] at (0.5,14.25) {\begin{minipage}{6cm}\begin{flalign*} \mathbf{M}_{\sigma,ij}^{(e)}&= & \end{flalign*}\end{minipage}};
\hdgstar{15.5,14}{26}

\draw [mybox] (0.5,12) rectangle (15.5,10);
\node [anchor=west] at (0.5,11.25) {\begin{minipage}{6cm}\begin{flalign*} \mathbf{f}_{\mathrm{s},i}^{(e)}&= & \end{flalign*}\end{minipage}};
\hdgstar{15.5,11}{27}

\node [anchor=north west] at (0.5,9) {\begin{minipage}{15cm}\begin{flushleft}%
  \begin{itemize}[nolistsep]
  \item In the case of triangular elements, two 3-by-3 matrices and one 3-by-1 vector are found for each element.
  \item The assembling process is the procedure bringing such local contributions together in an overall algebraic system of equations.
  \end{itemize}
  \end{flushleft}\end{minipage}};

\draw [mybox] (0.5,4) rectangle (15.5,0.5);
\node [anchor=north west] at (0.5,4) {\begin{minipage}{15cm}\begin{flushleft}%
  \vspace{0.5cm}\begin{itemize}[nolistsep]
  \item Think how assembling can be organised in an efficient way when using Matlab.
  \item Implement these formulae in \texttt{curlcurl\_ll.m}, \texttt{edgemass\_ll.m} and \texttt{current\_Pstr.m}.
  \end{itemize}
  \end{flushleft}\end{minipage}};
\hdgstar{15.5,2}{28}
\end{tikzpicture}%
  \caption{Worksheet for calculating the coefficients of the algebraic system of equations. Here, a bit of calculus is required to come up with expressions for the matrix coefficients and right-hand-side contributions. Task \protect\tikz[baseline=-0.5ex]{
      \protect\node[star,star points=7,star point ratio=0.6,draw=black,fill=black!30] at (0,0) {28};
  } requires a first implementation action at the implementation points in three routines and thereby marks the start of the third part of the exercise. Here, the students may need some support when coding the first lines.}
  \label{fig:sh19_computecoefficients2}
\end{figure}
\begin{figure}[t]
  \centering
  \begin{tikzpicture}
\Large
\draw[rounded corners] (0,\hdgbottom) rectangle (16,20);
\node[anchor=west, fancytitle, rounded corners] at (2,20) {\textbf{Magnetic Flux Density}};

\node [anchor=north west] at (0.5,19) {\begin{minipage}{15cm}\begin{flushleft}%
  \begin{itemize}
  \item The magnetic flux density is
  \begin{flalign*}
    \hspace{1cm}B_x &=\frac{\partial A_z}{\partial y} &\\
    B_y &=-\frac{\partial A_z}{\partial x}
  \end{flalign*}
  \item The magnetic flux density in element $e$ is expressed in function of the degrees of freedom $\mathbf{u}$ and the shape functions by
  \end{itemize}
  \end{flushleft}\end{minipage}};

\draw [mybox] (2.5,13) rectangle (15.5,11);
\node [anchor=west] at (2.5,12.25) {\begin{minipage}{6cm}\begin{flalign*} B_x&= & \end{flalign*}\end{minipage}};

\draw [mybox] (2.5,10.5) rectangle (15.5,8.5);
\node [anchor=west] at (2.5,9.75) {\begin{minipage}{6cm}\begin{flalign*} B_y&= & \end{flalign*}\end{minipage}};

\node [anchor=north west] at (0.5,7.5) {\begin{minipage}{15cm}\begin{flushleft}%
  \normalsize Remark:	Notice the inversion of coordinates and the minus sign in the above formulae. This is typical for the curl-curl case. In case of an electrostatic formulation in terms of the electric scalar potentials, the formulae for $E_x$ and $E_y$ would look completely different.
  \end{flushleft}\end{minipage}};

\node [anchor=north west] at (2,5.5) {\begin{minipage}{12.5cm}\begin{flushleft}%
  \begin{itemize}[nolistsep]
  \item Derive the coefficients for $B_x$ and $B_y$.
  \item Implement these in \texttt{curl.m}.
  \item Compute the element-wise magnitude of the magnetic flux density (use the function \texttt{pyth.m}).
  \item Search for the element with the highest magnetic flux density.
  \item Compute the magnetic energy relying upon these values for the magnetic flux density.
  \end{itemize}
  \end{flushleft}\end{minipage}};
\hdgstar{1.5,3.5}{29}

\end{tikzpicture}%
  \caption{Worksheet for calculating the magnetic flux density from the FE solution for the magnetic vector potential.}
  \label{fig:sh20_magnfluxdens}
\end{figure}
\begin{figure}[t]
  \centering
  \begin{tikzpicture}
\Large
\draw[rounded corners] (0,\hdgbottom) rectangle (16,20);
\node[anchor=west, fancytitle, rounded corners] at (2,20) {\textbf{Magnetic Energy}};

\node [anchor=north west] at (0.5,19) {\begin{minipage}{13.5cm}\begin{flushleft}%
  The magnetic energy equals (only in the linear case)
  \begin{flalign*}
    \hspace{1cm}W_\text{magn} &=\frac{1}{2}\mathbf{u}^T\mathbf{K}_\nu\mathbf{u} \,,&
  \end{flalign*}
  where $\mathbf{u}$ is the vector of degrees of freedom and $\mathbf{K}_\nu$ is the curl-reluctance-curl matrix.  \end{flushleft}\end{minipage}};

\node [anchor=north west] at (2,15) {\begin{minipage}{11.5cm}\begin{flushleft}%
  \begin{itemize}[nolistsep]
  \item Post-process for the magnetic energy using FEMM
  \item \vspace{2.5cm} Compute the magnetic energy based on the FEMM solution and the reluctance matrix computed by your own.
  \item[] {\normalsize (the FEMM solution is the third column in \texttt{prb.node})}
  \item[] {\normalsize (the values in \texttt{prb.node} are line-integrated magnetic vector potentials (in Wb), despite of the fact that FEMM uses and stores magnetic vector potentials (in Tm) in the \texttt{.ans} file.)}
  \end{itemize}
  \end{flushleft}\end{minipage}};
\hdgstar{1.5,14}{30}
\hdgstar{1.5,10}{31}

\draw [mybox] (3.5,14) rectangle (15.5,12);
\node [anchor=west] at (3.5,13.25) {\begin{minipage}{6cm}\begin{flalign*} W_\text{magn}&= & \end{flalign*}\end{minipage}};

\node [anchor=north west] at (0.5,7) {\begin{minipage}{13.5cm}\begin{flushleft}%
  The magnetic energy also equals (only in the linear case)
  \begin{flalign*}
  \hspace{1cm}W_\text{magn} &=\frac{1}{2}LI^2 \,,&
  \end{flalign*}
  where $I$ is the the applied current and $L$ is the inductance
  \end{flushleft}\end{minipage}};

\node [anchor=north west] at (2,3.5) {\begin{minipage}{11.5cm}\begin{flushleft}%
  \begin{itemize}
  \item Compute the inductance of the magnet.
  \end{itemize}
  \end{flushleft}\end{minipage}};
\hdgstar{1.5,2.5}{32}

\draw [mybox] (3.5,2.5) rectangle (15.5,0.5);
\node [anchor=west] at (3.5,1.75) {\begin{minipage}{6cm}\begin{flalign*} L&= & \end{flalign*}\end{minipage}};

\end{tikzpicture}%
  \caption{Worksheet for calculating the magnetic energy and the inductance from the FE solution. The magnetic energy $W_\text{magn}$ can be calculated using the own implementation for $\mathbf{K}_\nu$ combined with the solution $\mathbf{u}$ already obtained by \texttt{FEMM}. This allows an early check of the own implementation, i.e., before boundary conditions are applied and the own system of equations is solved. The values for $W_\text{magn}$ and for the inductance $L$ can be compared with the results previously obtained in \texttt{FEMM}.}
  \label{fig:sh21_magnenergy}
\end{figure}
\begin{figure}[t]
  \centering
  \begin{tikzpicture}
\Large
\draw[rounded corners] (0,\hdgbottom) rectangle (16,20);
\node[anchor=west, fancytitle, rounded corners] at (2,20) {\textbf{Boundary Conditions}};

\node [anchor=north west] at (0.5,19) {\begin{minipage}{15cm}\begin{flushleft}%
  \begin{itemize}[nolistsep]
  \item The only boundary conditions present are homogeneous Dirichlet boundary conditions.
  \item Unconstrained nodes (subscript \texttt{b}, index set \texttt{idxdof}) are distinguished from constrained nodes (subscript \texttt{c}, index set \texttt{idxdir}).
  \item The unconstrained system would be
  \begin{flalign*}
    \hspace{1cm}\left[\begin{array}{cc} \mathbf{K}_\text{bb} & \mathbf{K}_\text{bc} \\ \mathbf{K}_\text{cb} & \mathbf{K}_\text{cc} \end{array}\right]
    \left[\begin{array}{c} \mathbf{u}_\text{b}  \\ \mathbf{u}_\text{c} \end{array}\right]
    &=\left[\begin{array}{c} \mathbf{f}_\text{b}  \\ \mathbf{f}_\text{c} \end{array}\right] \,.&
  \end{flalign*}
  \item Adding constraints leads to ($\mathbf{I}_\text{cq}\mathbf{y}_\text{b}$ denotes the boundary-integral term)
  \begin{flalign*}
    \hspace{1cm}\left[\begin{array}{ccc} \mathbf{K}_\text{bb} & \mathbf{K}_\text{bc} & 0 \\ \mathbf{K}_\text{cb} & \mathbf{K}_\text{cc} & \mathbf{I}_\text{cq} \\
      0 & \mathbf{I}_\text{qc} & 0 \end{array}\right]
    \left[\begin{array}{c} \mathbf{u}_\text{b}  \\ \mathbf{u}_\text{c} \\ \mathbf{y}_\text{q} \end{array}\right]
    &=\left[\begin{array}{c} \mathbf{f}_\text{b}  \\ \mathbf{f}_\text{c} \\ \mathbf{u}_\text{c} \end{array}\right] \,.&
  \end{flalign*}
  \item The values for $\mathbf{u}_\text{c}$ are known. Shifting $\mathbf{u}_\text{c}$ to the right-hand side and eliminating the Lagrange multipliers $\mathbf{y}_\text{q}$ leads to the constrained system $\mathbf{K}_\text{bb}\mathbf{u}_\text{b}=\mathbf{f}_\text{b}-\mathbf{K}_\text{bc}\mathbf{u}_\text{c}$.
  \end{itemize}
  \end{flushleft}\end{minipage}};

\node [anchor=north west] at (2,8) {\begin{minipage}{11.5cm}\begin{flushleft}%
  \begin{itemize}[nolistsep]
  \item ''Shrink'' the unconstrained system up to the constrained system.
  \item Solve the system of equations.
  \item ''Inflate'' the solution vector to a full solution vector including the constrained nodes.
  \item Compute the magnetic energy and compare to previously obtained values.
  \item Write the solution to a FEMM \texttt{.ans} file.
  \item {\small (be aware of the fact that we work with line-integrated magnetic vector potentials (Wb), whereas FEMM uses magnetic vector potentials (Tm), the \texttt{save\_femm} routine does the necessary conversion, have a look inside.)}
  \item Plot your own solution using FEMM.
  \end{itemize}
  \end{flushleft}\end{minipage}};
\hdgstar{1.5,5}{33}

\end{tikzpicture}%
  \caption{Worksheet for explanation the introduction of boundary conditions in the algebraic system of equations; procedure for solving the constrained system of equations. The introduction of boundary conditions is a comparably difficult part of a FE solver. For that purpose, two routines, i.e. \texttt{shrink} and \texttt{inflate}, hiding all technicalities, are provided. Nevertheless, the worksheet explains the impact of Dirichlet boundary conditions on the algebraic system of equations and explicitly points the students to the lines in the code where the boundary conditions are inserted.}
  \label{fig:sh22_boundaryconditions}
\end{figure}
\begin{figure}[t]
  \centering
  \begin{tikzpicture}
\Large
\draw[rounded corners] (0,\hdgbottom) rectangle (16,20);
\node[anchor=west, fancytitle, rounded corners] at (2,20) {\textbf{Aperture Field (1/2)}};

\node [anchor=north west] at (0,19) {\begin{minipage}{15.5cm}\begin{flushleft}%
  \begin{itemize}[nolistsep]
  \item The magnetic vector potential inside a circle with reference radius $r_\text{ref}$ in the aperture can be expressed by
  \begin{flalign*}
    \hspace{1cm}A_z(r,\varphi) &=\sum_{p\in P} \left(a_p\cos(p\varphi) +b_p\sin(p\varphi)\right) \left(\frac{r}{r_\text{ref}}\right)^p \,.&
  \end{flalign*}
  \item Then, the magnetic flux density is
  \begin{flalign*}
    \hspace{1cm}B_x(r,\varphi) &=\sum_{p\in P_0} \frac{p}{r}\left(-a_p\sin((p-1)\varphi) +b_p\cos((p-1)\varphi)\right) \left(\frac{r}{r_\text{ref}}\right)^p ;&\\
    B_y(r,\varphi) &=\sum_{p\in P_0} \frac{p}{r}\left(-a_p\cos((p-1)\varphi) -b_p\sin((p-1)\varphi)\right) \left(\frac{r}{r_\text{ref}}\right)^p .&
  \end{flalign*}
  \item The Fourier coefficients of the magnetic flux density evaluated at the reference circle are
  \begin{flalign*}
    \hspace{1cm}\mathcal{F}(B_x) &=\left(\frac{p}{r_\text{ref}}b_p,-\frac{p}{r_\text{ref}}a_p\right) &\\
    \mathcal{F}(B_y) &=\left(\hspace{0.3cm}-\frac{p}{r_\text{ref}}a_p\hspace{0.3cm},\hspace{0.3cm}-\frac{p}{r_\text{ref}}b_p\hspace{0.3cm}\right) &
  \end{flalign*}
  \item \vspace{1cm}These coefficients are called \\
  \hspace{1cm}\emph{harmonic components}\\
  \hspace{1cm}\emph{skew harmonic components}\\
  (here under the assumption of a vertical main dipole field).
  \end{itemize}
  \end{flushleft}\end{minipage}};

\draw[rounded corners] (4.8,8.3) rectangle (6.6,6.8);
\draw[rounded corners] (7.0,8.3) rectangle (9.0,6.8);
\draw plot [smooth] coordinates {(5.6,6.8) (5.6,6) (7.3,6) (7.3,4.8) (6.8,4.8)};
\draw plot [smooth] coordinates {(8,6.8) (8,6.2) (9.5,6.2) (9.5,4.2) (7.8,4.2)};

\end{tikzpicture}%
  \caption{Worksheet for explaining the concept of harmonic components and skew-harmonic components of the~aperture field of an accelerator magnet.}
  \label{fig:sh23_aperturefield1}
\end{figure}
\begin{figure}[t]
  \centering
  \begin{tikzpicture}
\Large
\draw[rounded corners] (0,\hdgbottom) rectangle (16,20);
\node[anchor=west, fancytitle, rounded corners] at (2,20) {\textbf{Aperture Field (2/2)}};

\node [anchor=north west] at (2,19) {\begin{minipage}{13.7cm}\begin{flushleft}%
  \begin{itemize}[nolistsep]
  \item Find the nodes in the mesh which lie on the reference circle (use a geometric tolerance!) and order them along the circle.
  \item Extract the magnetic vector potential at the quarter of the reference circle which is inside the computational domain (notice: our solution consists of line-integrated magnetic vector potentials!).
  \item\vspace{0.5cm} Combine this signal with itself until you have a periodic signal.
  \item Make a Fourier transformation to obtain the coefficients for $A_z$.
  \item Compute the harmonic and skew harmonic components.
  \item\vspace{0.5cm} Make bar plots of both, discard the main dipole component and plot again.
  \item Fill the most important components in the table.
  \end{itemize}
  \end{flushleft}\end{minipage}};
\hdgstar{1.5,17.0}{34}
\hdgstar{1.5,13.5}{35}
\hdgstar{1.5,11.0}{36}

\draw[draw=black, fill=black!10] (0.5,9.0) rectangle (3.5,7.5);
\draw[draw=black, fill=black!10] (0.5,7.5) rectangle (3.5,6); \node [anchor=west] at (0.5,6.75) {dipole};
\draw[draw=black, fill=black!10] (0.5,6) rectangle (3.5,4.5); \node [anchor=west] at (0.5,5.25) {quadrupole};
\draw[draw=black, fill=black!10] (0.5,4.5) rectangle (3.5,3.0); \node [anchor=west] at (0.5,3.75) {sextupole};
\draw[draw=black, fill=black!10] (3.5,9) rectangle (9.5,7.5); \node [anchor=west] at (3.5,8.25) {harmonic component};
\draw[draw=black, fill=black!10] (3.5,7.5) rectangle (9.5,6);
\draw[draw=black, fill=black!10] (3.5,6) rectangle (9.5,4.5);
\draw[draw=black, fill=black!10] (3.5,4.5) rectangle (9.5,3);
\draw[draw=black, fill=black!10] (9.5,9) rectangle (15.5,7.5); \node [anchor=west] at (9.5,8.25) {skew harmonic component};
\draw[draw=black, fill=black!10] (9.5,7.5) rectangle (15.5,6);
\draw[draw=black, fill=black!10] (9.5,6) rectangle (15.5,4.5);
\draw[draw=black, fill=black!10] (9.5,4.5) rectangle (15.5,3);
\draw[draw=black, very thick] (0.5,9) rectangle (15.5,3);

\node [anchor=north west] at (0.5,2.8) {\begin{minipage}{15.2cm}\begin{flushleft}%
  The quality of the aperture field of an accelerator magnet is determined by the ratio of the magnitude of the higher harmonic components with respect to the main component (in this case the vertical dipole component). In typical magnets, this ratio is below $10^{-4}$. What about this magnet?
  \end{flushleft}\end{minipage}};

\end{tikzpicture}%
  \caption{Procedure for calculating the (skew) harmonic components of the SIS-100 magnet and worksheet for tabulating the results. This part requires a considerable amount of implementation but has a high relevance for the example of an accelerator magnet. At this point, a student group can decide to split up where a first subgroup addresses the aperture field, whereas a second subgroup proceeds with nonlinear materials.}
  \label{fig:sh24_aperturefield2}
\end{figure}
\begin{figure}[t]
  \centering
  \input{degersem2_sheet24}
  \caption{Definition of quantities involved with the modelling of nonlinear materials. The worksheet introduces the~notions of \emph{chord} and \emph{differential} reluctivity and puts them into relation to each other. The concept of an~\emph{operating point} at the BH-characteristic is explained. It should become clear to the students that each element of the iron-yoke part of the FE model may be operated at a different nonlinear operating point.}
  \label{fig:sh25_nonlinearmaterialproperties}
\end{figure}
\begin{figure}[t]
  \centering
  \begin{tikzpicture}
\Large
\draw[rounded corners] (0,\hdgbottom) rectangle (16,20);
\node[anchor=west, fancytitle, rounded corners] at (2,20) {\textbf{Successive Substitution}};

\node [anchor=north west] at (0.5,19) {\begin{minipage}{15cm}\begin{flushleft}%
  \begin{flalign*}
    \text{algorithm:}\quad \nabla\times\left(\nu(\textcolor{blue}{\vec{A}_n})\nabla\times\textcolor{red}{\vec{A}_{n+1}^*}\right) &=\vec{J}_\text{s} \,.&
   \end{flalign*}
   {\normalsize The convergence of a successive-substitution approach is poor. For that reason, commonly, relaxation is applied. We will use relaxation with a fixed relaxation factor.}
   \begin{flalign*}
    \text{relaxation:}\quad \textcolor{blue}{\vec{A}_{n+1}} &=\alpha\textcolor{red}{\vec{A}_{n+1}^*} +(1-\alpha)\textcolor{blue}{\vec{A}_n}) \,,&\\
   \end{flalign*}
   with relaxation factor $\alpha$.\\
   \vspace{0.5cm}{\normalsize The convergence of the nonlinear iteration is checked on the basis of a convergence criterion. Monitoring the convergence of the magnetic energy is the most appropriate. Here, we apply a simpler criterion, based on the relative difference between two successively obtained solutions.}
   \begin{flalign*}
      \text{convergence criterion:}\quad \varepsilon_\text{nlin} &=\frac{\left\|\textcolor{blue}{\vec{A}_{n+1}}\right\| -\left\|\textcolor{blue}{\vec{A}_n}\right\|}{\left\|\textcolor{blue}{\vec{A}_n}\right\|} \,.&
   \end{flalign*}   
   \end{flushleft}\end{minipage}};

\node [anchor=north west] at (2,6) {\begin{minipage}{13.5cm}\begin{flushleft}%
  \begin{itemize}[nolistsep]
  \item Implement the successive-substitution approach with relaxation and a convergence check.
  \item Solve the nonlinear problem and compare to results obtained with FEMM.
  \item What are the main differences between the linear and the nonlinear solution.
  \end{itemize}
  \end{flushleft}\end{minipage}};
\hdgstar{1.5,4}{38}

\end{tikzpicture}%
  \caption{Worksheet for explaining and implementing the successive-substitution method, combined with a simple relaxation technique. The students are invited to implement the successive-substitution method in the main file of the software.}
  \label{fig:sh26_successivesubstitution}
\end{figure}
\begin{figure}[t]
  \centering
  \begin{tikzpicture}
\Large
\draw[rounded corners] (0,\hdgbottom) rectangle (16,20);
\node[anchor=west, fancytitle, rounded corners] at (2,20) {\textbf{Newton (1/2)}};

\node [anchor=north west] at (0.5,19) {\begin{minipage}{15cm}\begin{flushleft}%
  \begin{flalign*}
    \text{algorithm 1:}\quad \nabla\times\left(\overline{\overline{\nu}}_\text{d}(\textcolor{blue}{\vec{A}_n})\nabla\times\delta\textcolor{red}{\vec{A}_{n+1}}\right) &=\vec{J}_\text{s}-\nabla\times\left(\nu(\textcolor{blue}{\vec{A}_n})\nabla\times\textcolor{blue}{\vec{A}_n}\right) \,.&
  \end{flalign*}
  The above form is commonly used. However, it may be more convenient to think about the Newton method as being similar to the successive-substitution approach, only differing concerning the linearisation of the working point. By introducing
  \begin{flalign*}\hspace{1cm}  \textcolor{red}{\vec{H}}&=\vec{H}_\text{m}(\textcolor{blue}{\vec{A}_n}) +\overline{\overline{\nu}}_\text{d}(\textcolor{blue}{\vec{A}_n})\textcolor{red}{\vec{B}} &
  \end{flalign*}
  in the magnetostatic equation, we arrive at
  \begin{flalign*}
    \text{algorithm 2:}\quad \nabla\times\left(\overline{\overline{\nu}}_\text{d}(\textcolor{blue}{\vec{A}_n})\nabla\times\textcolor{red}{\vec{A}_{n+1}}\right) &=\vec{J}_\text{s}-\nabla\times\vec{H}_\text{m}(\textcolor{blue}{\vec{A}_n}) \,.&
  \end{flalign*}
  Here, no increments are needed and only a matrix assembly is necessary for the left side of the equation.
Notice that the differential reluctivity is a tensor ($2$-by-$2$ in the 2D case, $3$-by-$3$ in the 3D case)!
  
  \vspace{1.3cm}The main challenge is the computation of the differential reluctivity tensor. In an element $e$ where the previous solution for the magnetic flux density is given by
  \begin{flalign*}
    \hspace{1cm}\textcolor{blue}{B^{(e)}} &=\left[\begin{array}{c}
      \textcolor{blue}{B_x^{(e)}} \\ \textcolor{blue}{B_y^{(e)}}
    \end{array}\right] \,,&
  \end{flalign*}
  the differential reluctivity tensor is
  \begin{flalign*}
    \hspace{1cm}\overline{\overline{\nu}}_\text{d}^{(e)} &=\nu^{(e)}\overline{\overline{1}} +2\textcolor{blue}{B^{(e)}}\nu_\text{d}^{(e)}\textcolor{blue}{B^{(e)T}} \,,&
  \end{flalign*}
  where $\nu^{(e)}$ and $\nu_\text{d}^{(e)}$ are the chord and differential reluctivities obtained by evaluating the material characteristic with input $\textcolor{blue}{B^{(e)}}$.
  \end{flushleft}\end{minipage}};

\end{tikzpicture}%
  \caption{Worksheet for explaining the Newton method. The Newton method is formulated as an alternative linearization method such that algorithm~2 is of the same form as in the linear case.}
  \label{fig:sh27_newton1}
\end{figure}
\begin{figure}[t]
  \centering
  \begin{tikzpicture}
\Large
\draw[rounded corners] (0,\hdgbottom) rectangle (16,20);
\node[anchor=west, fancytitle, rounded corners] at (2,20) {\textbf{Newton (2/2)}};

\node [anchor=north west] at (2,19) {\begin{minipage}{13.5cm}\begin{flushleft}%
  \begin{itemize}[nolistsep]
  \item Implement the nonlinear system matrix and additional right-hand-side term in \texttt{curlcurl\_ll\_nonlinear.m}.
  \item Set up a Newton iteration and check for convergence.
  \item Write the solution to a FEMM \texttt{.ans} file and compare with results obtained by FEMM.
  \item[] \vspace{0.5cm}{\normalsize Remark: The convergence of a Newton approach should be good enough to be convergent without relaxation.}
  \item \vspace{1cm}Compare the convergence of the successive-substitution approach with the convergence of the Newton approach. Try to find an optimal relaxation factor for the successive-substitution approach.
  \item \vspace{1cm}Compute the harmonic components and the skew harmonic components based on the nonlinear solution.
  \item Compare to the values obtained for the linear solution.
  \end{itemize}
  \end{flushleft}\end{minipage}};
\hdgstar{1.5,17.5}{39}
\hdgstar{1.5,12}{40}
\hdgstar{1.5,8.5}{41}

\draw[draw=black, fill=black!10] (0.5,6.5) rectangle (3.5,5);
\draw[draw=black, fill=black!10] (0.5,5) rectangle (3.5,3.5); \node [anchor=west] at (0.5,4.25) {dipole};
\draw[draw=black, fill=black!10] (0.5,3.5) rectangle (3.5,2); \node [anchor=west] at (0.5,2.75) {quadrupole};
\draw[draw=black, fill=black!10] (0.5,2) rectangle (3.5,0.5); \node [anchor=west] at (0.5,1.25) {sextupole};
\draw[draw=black, fill=black!10] (3.5,6.5) rectangle (9.5,5); \node [anchor=west] at (3.5,5.75) {harmonic component};
\draw[draw=black, fill=black!10] (3.5,5) rectangle (9.5,3.5);
\draw[draw=black, fill=black!10] (3.5,3.5) rectangle (9.5,2);
\draw[draw=black, fill=black!10] (3.5,2) rectangle (9.5,0.5);
\draw[draw=black, fill=black!10] (9.5,6.5) rectangle (15.5,5); \node [anchor=west] at (9.5,5.75) {skew harmonic component};
\draw[draw=black, fill=black!10] (9.5,5) rectangle (15.5,3.5);
\draw[draw=black, fill=black!10] (9.5,3.5) rectangle (15.5,2);
\draw[draw=black, fill=black!10] (9.5,2) rectangle (15.5,0.5);
\draw[draw=black, very thick] (0.5,6.5) rectangle (15.5,0.5);

\end{tikzpicture}%
  \caption{Worksheet for implementing the Newton method and calculating the (skew) harmonic components in the~magnet's aperture accounting for ferromagnetic saturation of the iron yoke. The implementation of the Newton method is quite involved. Nevertheless, the result obtained by the students themselves is a nonlinear magnetostatic FE solver which is from the algorithmic side close to optimal. The results for the (skew) harmonic components should be compared to the results obtained for the (virtual) linear case. One should notice that the higher-order components in the nonlinear case are considerably lower than the ones for the linear case, which indicates that the~magnet has been optimized for providing a possibly high homogeneity at maximal field.}
  \label{fig:sh28_newton2}
\end{figure}

\end{document}